\documentclass{article}
\usepackage[preprint]{neurips_2024}
\usepackage{tabularx}
\usepackage{booktabs}
\usepackage{graphicx}
\makeatletter
\renewcommand{\@noticestring}{\rule{2in}{0.4pt}\\$^*$Equal contribution.\\Preprint.}
\makeatother

\usepackage{amsmath}
\usepackage{microtype}
\usepackage{xcolor}
\usepackage{listings}
\usepackage{enumitem}
\usepackage[hyphens]{url}

\title{EnterpriseRAG-Bench: A RAG Benchmark\\for Company Internal Knowledge}

\author{
  Yuhong Sun$^{1*}$ \quad
  Joachim Rahmfeld$^{1*}$ \quad
  Chris Weaver$^{1}$ \quad
  Weijia Chen \\[0.3em]
  \bf Roshan Desai$^{1}$ \quad
  Wenxi Huang$^{1}$ \quad
  Mark H.\ Butler$^{2}$ \\[0.5em]
  $^{1}$Onyx \quad $^{2}$UC Berkeley \\[0.3em]
  \texttt{\{yuhong, joachim\}@onyx.app} 
}

\begin{document}

\maketitle

\begin{abstract}
Retrieval-Augmented Generation (RAG) has become the standard approach for grounding large language models in information that was not available during training. While existing datasets and benchmarks focus on web or other public sources, there is still no widely adopted dataset that realistically reflects the nature of company-internal knowledge. Meanwhile, startups, enterprises, and researchers are increasingly developing AI Agents designed to operate over exactly this kind of proprietary data. To close this gap, we release a synthetic enterprise corpus, its generation framework, and a leaderboard.

We present \textbf{EnterpriseRAG-Bench}, a dataset consisting of approximately \textbf{500,000 documents} spanning nine enterprise source types (Slack, Gmail, Linear, Google Drive, HubSpot, Fireflies, GitHub, Jira, and Confluence) and \textbf{500 questions} across ten categories that test distinct retrieval and reasoning capabilities. The corpus is generated with cross-document coherence---grounded in shared projects, people, and initiatives---and augmented with realistic noise such as misfiled documents, near-duplicates, and conflicting information. The question set ranges from simple single-document lookups to multi-document reasoning, constrained retrieval, conflict resolution, and recognizing when information is absent. The generation framework lets teams generate variants tailored to their own industry, scale, and source mix. The dataset, code, evaluation harness, and leaderboard are available at \url{https://github.com/onyx-dot-app/EnterpriseRAG-Bench}.
\end{abstract}

\section{Introduction}
\label{sec:intro}

\begin{table}[t!]
  \caption{Enterprise source types and approximate document volumes.}
  \label{tab:sources}
  \centering
  \small
  \begin{tabular*}{\textwidth}{@{\extracolsep{\fill}}clrp{8.5cm}}
    \toprule
    \# & Source Type   & Approx.\ Volume & Description \\
    \midrule
    1  & Slack         & 275,000 & Internal channels and team discussions \\
    2  & Gmail         & 120,000 & Email threads from management, sales, leadership, and ICs \\
    3  & Linear        &  35,000 & Project management tickets for engineering, product, and design \\
    4  & Google Drive  &  25,000 & Shared files and collaborative documents \\
    5  & HubSpot       &  15,000 & CRM records for sales pipeline and customer tracking \\
    6  & Fireflies     &  10,000 & Meeting transcripts for internal and external-facing meetings \\
    7  & GitHub        &   8,000 & Pull requests and review comments across code repositories \\
    8  & Jira          &   6,000 & Support tickets, both internal and customer-facing \\
    9  & Confluence    &   5,000 & Wikis, runbooks, and structured documentation \\
    \bottomrule
  \end{tabular*}

  \vspace{1.2em}

  \caption{Question categories, counts, and targeted capabilities.}
  \label{tab:question_types}
  \small
  \begin{tabular*}{\textwidth}{@{\extracolsep{\fill}}clcp{8.5cm}}
    \toprule
    \# & Category & Count & Description \\
    \midrule
     1 & Basic                    & 175 & Single ground truth document; moderate keyword overlap. \\
     2 & Semantic                 & 125 & Roundabout phrasing with low keyword overlap; requires semantic matching. \\
     3 & Intra-Doc Reasoning      &  40 & Combining information from distant sections of one long document. \\
     4 & Project Related          &  40 & Aggregating knowledge from related documents within a single project. \\
     5 & Constrained              &  30 & Qualifiers narrow multiple relevant documents to one correct answer. \\
     6 & Conflicting Info         &  20 & Documents contradict each other; requires a complete, correct answer. \\
     7 & Completeness             &  20 & Requires fetching all relevant documents exhaustively. \\
     8 & Miscellaneous            &  20 & Targets informal, off-topic, or loosely organized documents. \\
     9 & High Level               &  10 & Synthesis across the corpus; no single gold document. \\
    10 & Info Not Found           &  20 & Answer is absent from the corpus; requires recognizing absence. \\
    \bottomrule
  \end{tabular*}
\end{table}

The growing adoption of Retrieval-Augmented Generation (RAG) has driven a wave of benchmark development for evaluating retrieval and generation quality. Prominent datasets such as Natural Questions~\citep{kwiatkowski2019natural}, MS~MARCO~\citep{bajaj2018msmarco}, and the BEIR meta-benchmark~\citep{thakur2021beir} have been instrumental in advancing the field, while multi-hop reasoning benchmarks like HotpotQA~\citep{yang2018hotpotqa}, MuSiQue~\citep{trivedi2022musique}, and BrowseComp-Plus~\citep{chen2025browsecomp} test more complex retrieval chains. Domain-specific efforts have also emerged, including PubMedQA~\citep{jin2019pubmedqa} for biomedical literature and FinQA~\citep{chen2021finqa} for financial reasoning. More recently, MTEB~\citep{muennighoff2023mteb} and KILT~\citep{petroni2021kilt} have provided unified evaluation frameworks spanning multiple tasks.

Despite this progress, these benchmarks share a common assumption: the underlying corpus is drawn from publicly accessible sources such as Wikipedia, discussion forums, web results, or scientific articles. Yet one of the most consequential applications of RAG today is over \textbf{company-internal knowledge} (an organization's documents, messages, tickets, and records), and no widely adopted public benchmark reflects its characteristics. For example, KARLBench~\citep{databricks2025karl} reflects growing industry interest in enterprise knowledge grounding, but its corpus is predominantly public documents with only a small unreleased proprietary subset.

Company-internal data differs from public corpora in ways that existing benchmarks do not capture. Real company data is disorganized, noisy, and spans document types absent from public benchmarks, such as support tickets, email threads, customer call transcripts, and chat conversations. To build a dataset that faithfully reflects these differences, we identify five design considerations:

\begin{enumerate}[leftmargin=2em]
  \item \textbf{Cross-document coherence.} Documents should be connected through shared initiatives, people, and decisions, not generated in isolation.
  \item \textbf{Realistic volume distribution.} The ratio of documents across source types and the topic distribution within each source should reflect real-world patterns.
  \item \textbf{Realistic noise.} The dataset should include misfiled documents, near-duplicates, and conflicting or outdated information rather than assuming a clean knowledge base.
  \item \textbf{Internal terminology.} Documents should use project codenames, product-specific acronyms, and organizational jargon that are essential for correct retrieval but meaningless outside the company.
  \item \textbf{Generality across enterprise settings.} The generation framework must support diverse industries, company stages, and organizational structures rather than encoding assumptions specific to a single domain.
\end{enumerate}

Beyond these design principles, there are also \textbf{pragmatic constraints} that any large-scale generation effort must address. Generating up to millions of documents with LLMs is expensive and time-consuming. Requiring full cross-document awareness for every document, while ideal for coherence, would make the generation process prohibitively costly. Our generation pipeline addresses this tension by using targeted strategies: high-fidelity, context-rich generation for smaller clusters of tightly coupled documents, and cost-efficient, scaffolding-guided generation for the high-volume bulk of the corpus.

\section{Dataset Overview}

The EnterpriseRAG-Bench corpus simulates a technology company called ``Redwood Inference'' and is organized across the source types listed in Table~\ref{tab:sources}. The question set spans ten categories (Table~\ref{tab:question_types}), each targeting a distinct retrieval or reasoning capability. The volume distribution is intentionally non-uniform: Slack and Gmail dominate the corpus by an order of magnitude over sources like Confluence, reflecting the reality that informal communication channels produce far more documents than curated knowledge bases.

Basic and Semantic questions test standard retrieval with varying levels of lexical overlap. Intra-Document Reasoning evaluates whether systems can synthesize information scattered across a long document rather than relying on a single passage. Project Related questions test multi-document reasoning over documents tied to a shared initiative. Constrained and Conflicting Info questions both require disambiguating among multiple plausibly-relevant documents: through narrowing qualifiers in the former, and contradictions in the latter. Completeness questions push further by demanding exhaustive recall of every relevant document. Miscellaneous questions target informal, off-topic, or ad-hoc documents stored in less predictable locations. High Level questions require synthesis across a broader portion of the corpus, where multiple distinct subsets of documents may each produce a complete and correct answer. Info Not Found questions test whether the system can recognize when the corpus does not contain the answer, rather than hallucinating one from superficially related documents.

\section{Dataset Analysis}

This section characterizes the structural patterns of EnterpriseRAG-Bench by comparing it against two reference corpora. The Onyx subset is drawn from the company behind this work, restricted to a subset of source types that overlap with the benchmark. Onyx, like the hypothetical Redwood Inference, is an AI software company, making it a close domain match for the comparison.\footnote{Future work will extend this comparison with other larger real-world enterprise datasets.} BrowseComp-Plus~\citep{chen2025browsecomp} is drawn from the open web and covers a broad, heterogeneous mix of topics, serving as an external baseline for comparison. The analysis indicates that EnterpriseRAG-Bench is structurally closer to the Onyx data than to BrowseComp-Plus. All embeddings are produced with OpenAI's \texttt{text-embedding-3-large}~\citep{openai2024embeddings}. Unless stated otherwise, documents are chunked into 512-token segments using the embedding model's tokenizer (\texttt{cl100k\_base} via \texttt{tiktoken}).

\subsection{A Visual Illustration}

\begin{figure}[t!]
    \centering
    \includegraphics[width=\textwidth]{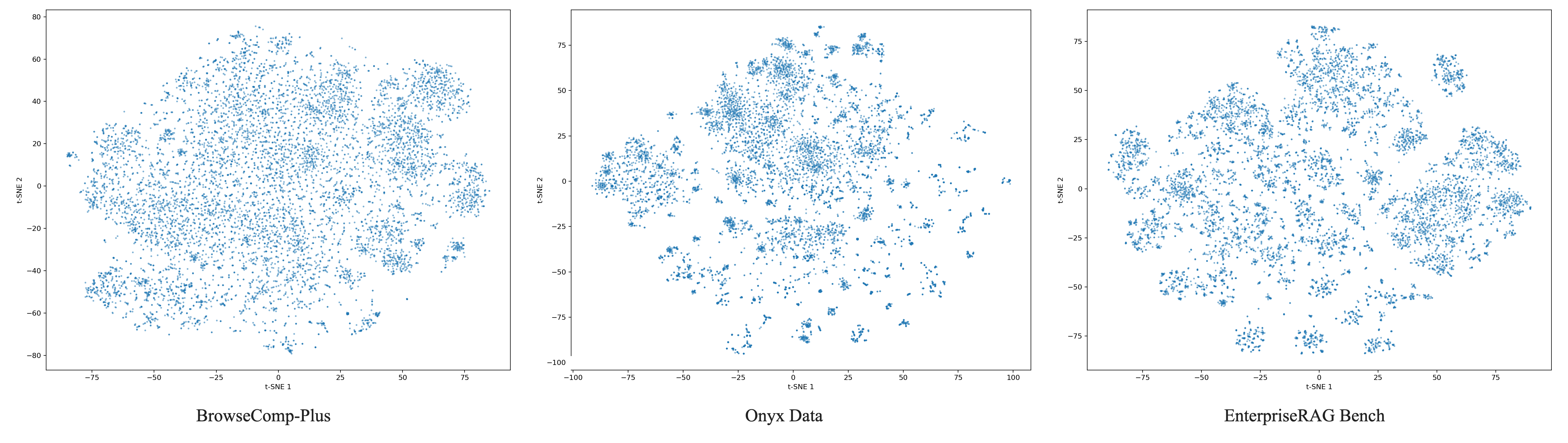}
    \caption{t-SNE projections for BrowseComp-Plus, Onyx data, EnterpriseRAG-Bench.}
    \label{fig:t-sne}
\end{figure}

Figure~\ref{fig:t-sne} shows t-SNE projections~\citep{vandermaaten2008tsne} of the three datasets, each computed from a sample of 10{,}000 document embeddings.

BrowseComp-Plus appears as a single broad, diffuse cloud with no pronounced sub-structure, as expected for an open-domain corpus. The Onyx data decomposes into clusters of varying sizes and densities, with some topic regions connected by thinner bridging areas where documents span multiple topics. EnterpriseRAG-Bench reproduces both the clustered structure and some of this bridging behavior, but its clusters appear more uniform---closer in size and tightness to one another than those in the Onyx data. We attribute this more predictable spread around the clusters to the topic-based scaffolding used to generate the synthetic dataset (further discussed in Section~\ref{sec:limitations}). Because t-SNE distorts distances, these observations are suggestive rather than conclusive, and motivate the quantitative metrics in the next subsection.

\subsection{Quantitative Analysis}
\label{sec:quant}

\begin{table}[t!]
  \caption{Average cosine similarity at local, in-cluster, and cross-cluster scales across the three corpora (k-means $k=10$; chunk size 8192).}
  \label{tab:acs_benchmark_comparison}
  \centering
  \small
  \begin{tabular*}{\textwidth}{@{\extracolsep{\fill}}lccc}
    \toprule
    \textbf{Avg. Cosine Similarity} & \textbf{EnterpriseRAG-Bench} & \textbf{Onyx company data} & \textbf{BrowseComp-Plus} \\
    \midrule
    Local (top 10 KNN) & 0.83 & 0.83 & 0.69 \\
    In-cluster         & 0.61 & 0.56 & 0.29 \\
    Cross-cluster      & 0.50 & 0.36 & 0.20 \\
    \bottomrule
  \end{tabular*}
\end{table}

We characterize each corpus with metrics that capture both local and global properties.

\textbf{Local density.} To quantify the density of each document's immediate semantic neighborhood, we draw a uniform random sample of 1{,}000 documents from each corpus. For each sampled document, we compute the average cosine similarity to its ten nearest neighbors in the embedding space; Table~\ref{tab:acs_benchmark_comparison} reports the mean of these values across the 1{,}000 samples. We use exhaustive pairwise comparison rather than ANN search~\citep{malkov2020hnsw} to avoid recall misses and ranking noise.

We rely on a dense embedding model for this measurement because the model is trained so that proximity in its output space reflects semantic similarity. A higher average cosine similarity therefore corresponds to a denser local neighborhood, where retrieval should be more challenging because many documents lie close to the target conceptually (see Figure~\ref{fig:recall-acs} in Section~\ref{sec:scale} for how this metric scales with subsets of the corpus).

\textbf{Global clustering.} We partition each corpus with k-means~\citep{lloyd1982kmeans} on the document embeddings and report two complementary statistics. \emph{In-cluster similarity} is the average cosine similarity between random pairs of documents that fall within the same cluster, indicating how tightly each cluster concentrates around its centroid. \emph{Cross-cluster similarity} is the same quantity for random pairs drawn from different clusters, indicating how well separated the clusters are from one another in semantic space.

The numbers in Table~\ref{tab:acs_benchmark_comparison} are reported for $k=10$, i.e., k-means partitions the corpus into ten clusters. We varied $k$ across a range of values from 5 through 50 and observed that, while absolute similarities shift modestly with $k$, the relative ordering and gaps between the three corpora are preserved; we therefore report a single representative setting. Each statistic is computed as the mean cosine similarity over 1{,}000 random document pairs drawn either within the same cluster or across distinct clusters.

\textbf{Interpretation.} EnterpriseRAG-Bench and the Onyx corpus exhibit nearly identical top-10 local similarities, both substantially higher than BrowseComp-Plus. This is a primary marker of retrieval difficulty. When a document's nearest neighbors lie close in embedding space, distractors are abundant, and the retriever cannot rely on a wide margin to separate the gold document from the rest of the corpus.

At the global level, both enterprise corpora show elevated in-cluster similarity, as enterprise documents are concentrated around a shared and recurring set of business contexts. Cross-cluster similarity is also elevated, since the overall span of topics is narrower than in an open-domain corpus. The gap between in-cluster and cross-cluster similarity reflects how cleanly topics divide. Enterprise corpora are expected to exhibit a wider gap than open-domain corpora, where documents span a broader and less-concentrated range of subjects. Table~\ref{tab:acs_benchmark_comparison} reflects these expectations.

EnterpriseRAG-Bench has higher average in-cluster and cross-cluster cosine similarity than Onyx. We attribute this to the LLM's difficulty generating realistic peripheral and off-topic content, discussed further in Section~\ref{sec:limitations}. The gap does not affect the core retrieval challenge, since peripheral documents are easily discarded by most retrieval systems. Real corpora such as Onyx contain such peripheral material, which inflates document counts without contributing to retrieval difficulty; EnterpriseRAG-Bench therefore behaves like a larger corpus than its raw document count would suggest.

\section{Data Generation Methodology}

\begin{figure}[t!]
  \centering
  \includegraphics[width=\textwidth]{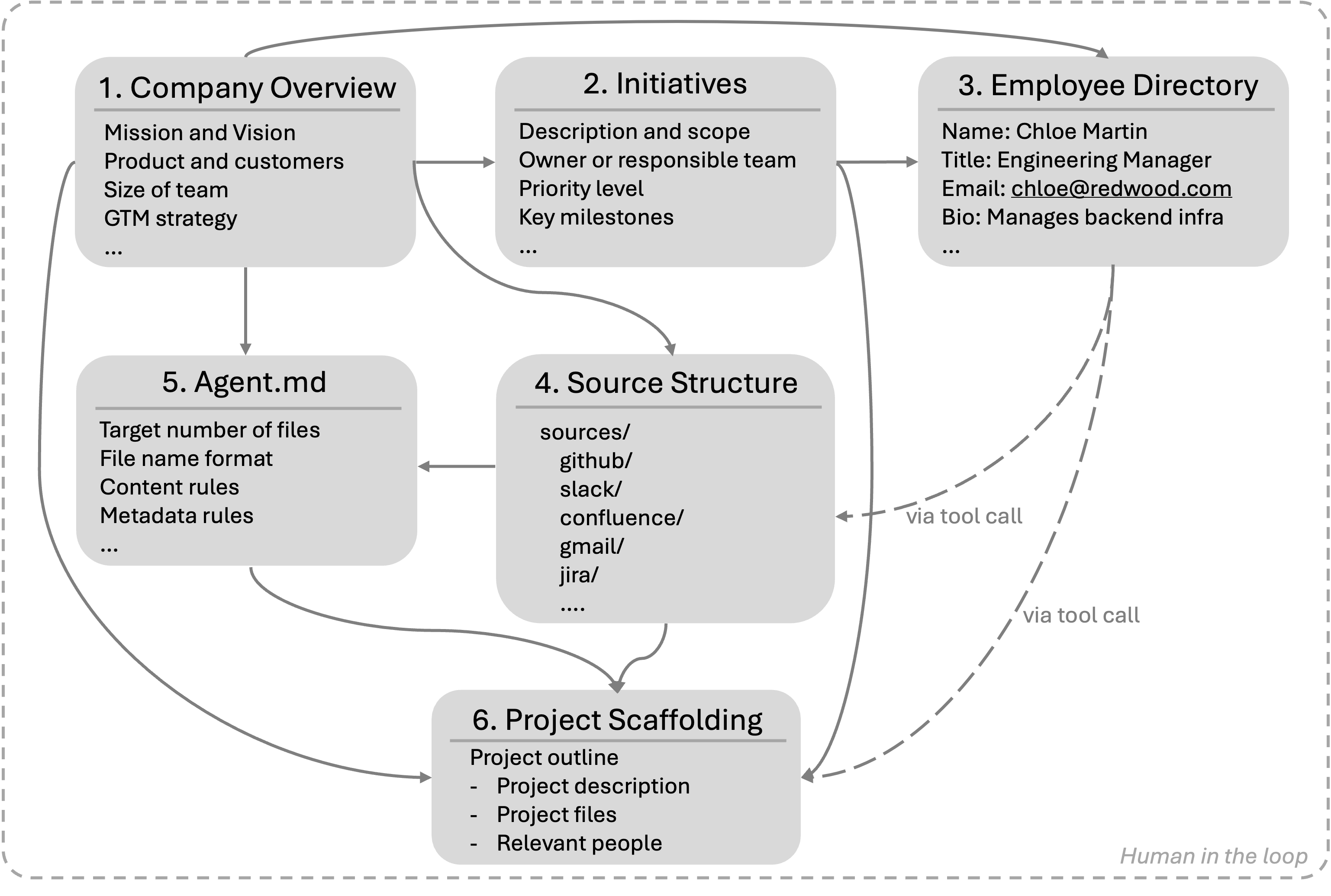}
  \caption{Overview of the generation framework scaffolding. Each downstream generation step is grounded in a subset of these five human-in-the-loop artifacts, ensuring cross-document coherence throughout the corpus.}
  \label{fig:generation_harness}
\end{figure}

The generation pipeline consists of three stages: building the document set, introducing noise, and generating questions. Two requirements shape the pipeline beyond the dataset-level design principles introduced in Section~\ref{sec:intro}. The corpus must be \textbf{question-aware}: certain question types---completeness, conflicting info, constrained---require specific document structures generated intentionally rather than discovered after the fact across hundreds of thousands of documents. And every question must have \textbf{verifiable ground truth}: gold documents and answers targeted enough for reliable evaluation without exhaustive manual review, paired with phrasing that avoids trivial keyword overlap that would let retrieval systems saturate the benchmark.

\subsection{Building the Document Set}
\label{sec:building}

Document generation proceeds in three phases: scaffolding, high-fidelity generation, and high-volume generation.

\paragraph{Scaffolding.}
All generation begins from five top-level artifacts produced through human-in-the-loop collaboration with an LLM (Figure~\ref{fig:generation_harness}):

\begin{enumerate}[leftmargin=2em]
  \item A \textbf{company overview} covering mission, products, technology, business model, competitive positioning, team size, and funding history.
  \item A set of \textbf{high-level initiatives} describing the company's major strategic efforts.
  \item An \textbf{employee directory} with names, roles, departments, and reporting relationships.
  \item A \textbf{source structure} defining the directory hierarchy for each of the nine source types, informed by the company context and real-world organizational patterns.
  \item \textbf{\texttt{agents.md} files} placed throughout the directory tree to specify, in natural language, the expected format and content of documents in each location (e.g., that GitHub contains only pull requests with review comments, not issues).
\end{enumerate}

These artifacts are referenced by every downstream generation step and ensure that documents share a common organizational reality. The scaffolding extends to the project level: initiatives are decomposed into projects, each targeting on the order of 100 documents, and individually enriched with detailed descriptions, document manifests, and assigned employees. Appendix~\ref{app:project_scaffolding} provides procedural details for the project-level decomposition.

\paragraph{High-fidelity generation.}
The core document set is produced with maximal cross-document awareness. Each document is generated with access to the company overview, its project's full manifest of sibling documents, the \texttt{agents.md} files along its path, and a read tool for inspecting previously generated project documents on demand. This ensures that documents within a project exhibit the implicit cross-references---shared decisions, referenced meetings, linked tickets---that characterize real enterprise knowledge. A separate step produces \textbf{completeness clusters}: groups of 4--10 documents generated with full mutual visibility and facts deliberately distributed across the group, supporting questions that require exhaustive recall.

\paragraph{High-volume generation.}
The bulk of the corpus is produced through a cost-efficient pipeline that trades per-document context for scale. The central challenge is \textbf{model drift}: without tight steering, the LLM converges on a narrow set of themes and produces near-duplicate documents. In a controlled experiment, we prompted the model to generate 100 documents using only the company overview for context. An LLM judge flagged over 40\% as having a very close sibling; the exact rate varied across generation models and temperature settings, but the pattern was always prevalent. To counteract this, we introduce a \textbf{topic scaffolding} layer that hierarchically decomposes each source type into topics and subtopics, with document counts calibrated to real-world distributions. Topics are recursively split until each leaf contains at most 500 documents. During generation, the model sees only the company overview, key initiatives, the relevant \texttt{agents.md}, and file paths within the same leaf topic. This narrow context window substantially reduces cost while eliminating the runaway duplication observed without scaffolding. Appendix~\ref{app:docgen} provides full procedural details for all three phases.

\subsection{Adding Noise}
\label{sec:noise}

Real enterprise corpora are not clean. After the base document set is generated, we introduce four types of noise to approximate real-world imperfections (Table~\ref{tab:noise}).

\begin{table}[t!]
  \caption{Noise types introduced after base document generation.}
  \label{tab:noise}
  \centering
  \small
  \begin{tabular*}{\textwidth}{@{\extracolsep{\fill}}lp{10cm}}
    \toprule
    Noise Type & Mechanism \\
    \midrule
    Random shuffle      & Documents relocated within their source type via directory-walk sampling \\
    LLM-based shuffle   & Documents relocated to structurally plausible but incorrect locations \\
    Near-duplicates     & New version generated with specific facts changed; may cross source types \\
    Miscellaneous files & Informal, off-topic content in ad-hoc directories (e.g., \texttt{slack/memes}, \texttt{google\_drive/.../misc-assets}) \\
    \bottomrule
  \end{tabular*}
\end{table}

Random shuffling samples documents via a directory walk rather than uniformly across the corpus, which avoids oversampling high-volume sources while leaving low-volume ones too clean. Shuffling is constrained to the same source type because the document is not rewritten: a GitHub pull request, for instance, cannot plausibly appear in a Fireflies sales-call directory. LLM-based shuffling operates similarly but relies on the model's understanding of the directory hierarchy to choose a less ideal but still plausible target location. The result is more realistic misfiling patterns, where documents end up in adjacent directories or nearby branches rather than arbitrary places. Near-duplicate pairs introduce controlled factual divergence and are tracked for later use in conflicting-info questions. Miscellaneous files support a question category targeting informal content in less predictable locations. Procedural details for each noise type are provided in Appendix~\ref{app:docgen}.

\subsection{Generating Questions}
\label{sec:questions}

The ten question types are designed to reflect the richness and complexity of realistic questions in enterprise settings. For example, projects that receive multiple updates over their lifetime give rise to questions about conflicting or outdated information, which is rarely emphasized in public corpora. Each type is produced through a distinct generation flow, and all flows share a common output structure: a question, a gold answer, a list of atomic answer facts for fine-grained completeness scoring, and (where applicable) gold document identifiers. The flows fall into three families based on their corpus-awareness requirements (Table~\ref{tab:qgen_families}).

\begin{table}[t!]
  \caption{Question generation families and their corpus-awareness approaches.}
  \label{tab:qgen_families}
  \centering
  \small
  \begin{tabular*}{\textwidth}{@{\extracolsep{\fill}}lp{3.8cm}p{6.2cm}}
    \toprule
    Family & Question Types & Approach \\
    \midrule
    Single-document      & Basic, Semantic, Intra-Doc Reasoning, Miscellaneous & Sample one document; generate a question with type-specific phrasing and validation constraints \\
    Corpus-exploration   & Project Related, Constrained, High Level, Info Not Found & LLM agent explores the corpus with file-system tools before formulating a question \\
    Generation-dependent  & Conflicting Info, Completeness & Built from document structures created in Sections~\ref{sec:building} and~\ref{sec:noise} \\
    \bottomrule
  \end{tabular*}
\end{table}

Single-document flows differ primarily in sampling criteria and prompt steering. For instance, semantic questions require roundabout phrasing with minimal lexical overlap, while intra-doc reasoning questions must link information from distant sections and are rejected if answerable from a single passage. Corpus-exploration flows equip the generation model with tools (glob, grep, read, etc.) to discover document relationships before proposing a question. Constrained questions additionally produce \textbf{anti-hallucination facts}: negative statements that catch errors a system might make if it retrieves a distractor document instead of the gold document. Generation-dependent flows leverage the near-duplicate pairs (Section~\ref{sec:noise}) and completeness clusters (Section~\ref{sec:building}) created during earlier stages, ensuring the document structures needed for these question types exist by construction. Detailed per-type generation flows are provided in Appendix~\ref{app:questions}.

\subsection{Quality Assurance}

Generating questions from documents does not guarantee that the source document is the only, or even the best, document for that question. A question back-generated from one document may be equally well or better answered by an unrelated document elsewhere in the corpus. Similarly, while project-level scaffolding enforces coherence within a project, a document produced during high-volume generation may coincidentally contain information that supplements or supersedes a project-based gold answer. At the scale of 500,000 documents, these collisions are not hypothetical edge cases but expected outcomes. Three mechanisms address this inherent uncertainty.

\paragraph{LLM-based conformance checks.}
Each question type includes a validation step that verifies type-specific requirements: for example, that intra-document reasoning questions cannot be answered from a single contiguous passage, or that high-level questions are not answerable from any one document. Questions failing validation are regenerated or discarded.

\paragraph{Multi-retriever pooling.}
Three independent retrieval systems, included in the released code, are run against the question set: a BM25 keyword search (OpenSearch), a vector search (OpenAI \texttt{text-embedding-3-large} embeddings with cosine similarity over Qdrant), and a bash agent with shell tools. Their retrieved documents are pooled with the original gold set and evaluated through the three-judge consensus process described in Section~\ref{sec:correction} to identify missing gold documents or flag incorrect ones. This pooling improves gold set coverage beyond what any single generation or retrieval pass can achieve, though it does not guarantee exhaustive correctness. Not all corrections are recoverable: for example, if pooling reveals a second gold document for a question type that requires exactly one, the question must be discarded rather than corrected, since updating the gold set would violate the type's structural constraints. To account for this loss, each question type was initially overgenerated beyond its target count. After conformance checks and retriever-based corrections, questions that received corrections were discarded and the remaining questions were randomly sampled down to the desired volume per category.

\paragraph{Ongoing human curation.}
It is more feasible to verify that a newly proposed document is better than the current gold document than to confirm exhaustively that no better match exists anywhere in the corpus. As benchmark submissions are collected, we will also collect corrections to gold documents and answers surfaced by participating systems. These corrections will be reviewed and periodically incorporated into updated releases of the dataset, allowing the ground truth to improve over time as more systems exercise different retrieval strategies against the dataset. We encourage the community to participate through either submissions to the leaderboard\footnote{\url{https://huggingface.co/spaces/onyx-dot-app/EnterpriseRAG-Bench-Leaderboard}} or issues on GitHub.\footnote{\url{https://github.com/onyx-dot-app/EnterpriseRAG-Bench}}

\section{Evaluation Framework}
\label{sec:evaluation}

Because exhaustive gold-label correctness is not feasible at this scale, the evaluation framework must treat gold labels as revisable hypotheses rather than fixed ground truth, while still producing stable, interpretable metrics. We address this through a correction-aware scoring pipeline (Section~\ref{sec:correction}) that updates gold labels when evidence warrants it, combined with metrics that separately measure answer quality and retrieval performance.

\subsection{Metrics}
\label{sec:metrics}

Each question is scored along four dimensions: correctness, completeness, document recall, and invalid extra documents.

\paragraph{Correctness} is a binary LLM-based judgment of whether the candidate answer is broadly aligned with the gold answer. The judge is lenient toward stylistic differences, additional context, and supplementary detail, but the answer must not contain factual conflicts with the gold answer or mismatched quantities.

\paragraph{Completeness} measures the fraction of atomic answer facts supported by the candidate answer. Each fact from the gold \texttt{answer\_facts} list is independently validated by an LLM judge that checks whether the candidate answer contains or implies that fact. The score is the fraction of facts validated. Because each fact is evaluated in isolation, no single judgment propagates error to others, and the metric provides fine-grained signal about which aspects of the answer are missing.

\paragraph{Document Recall} is the fraction of gold documents that appear in the candidate's retrieved set, computed only for question types that carry expected document identifiers. If not indicated otherwise, Recall@10 is implied.

\paragraph{Invalid Extra Documents} counts retrieved documents that are neither in the gold set nor classified as ``valid'' (relevant but not required) by the evaluation judges, again restricted to question types with expected documents. This metric captures retrieval noise: systems that retrieve broadly may achieve high recall at the cost of surfacing irrelevant material. We report an absolute count rather than precision because the answer generator is affected by the volume of irrelevant content it must process, not by its proportion of the retrieved set.

Correctness and completeness are evaluated independently: the correctness judge has no visibility into the answer facts, and fact validation has no visibility into the correctness judgment. Prior to scoring, citations are stripped from candidate answers to prevent the judge from being biased by reference formatting rather than substantive content.

\paragraph{Leaderboard.} The leaderboard ranks systems by a single aggregate score. For each question, the score is the completeness percentage if the answer is correct and zero otherwise. The overall score is the average across all questions.

\subsection{Comparative Evaluation}

For head-to-head system comparison, the framework supports a comparative mode that evaluates two systems on the same question set. Retrieved documents from both systems are pooled through the same three-judge correction process (Section~\ref{sec:correction}), and both systems are scored against the resulting (potentially updated) gold set. A separate three-judge panel produces a per-question preference between the two systems' answers; to mitigate positional bias, system ordering is randomly swapped before presentation and mapped back afterward. Retrieved documents are partitioned into shared and system-unique subsets to support analysis of where systems diverge.

\subsection{Gold Set Correction}
\label{sec:correction}

\begin{quote}
\textit{Note: All leaderboard submissions are scored against the same fixed gold set---the one included in the initial release---so results are directly comparable regardless of when a system is evaluated. Corrections are accumulated offline and incorporated into future versioned releases; no retroactive rescoring occurs within a release version.}
\end{quote}

A key design decision is that the gold set is not static. When a system retrieves documents that differ from the annotated gold set, the evaluation pipeline determines whether the gold labels should be updated before scoring. The correction procedure operates as follows.

The union of gold documents and candidate-retrieved documents is evaluated through a \textbf{three-judge consensus} process. Because document relevance can be ambiguous and because individual LLM judgments exhibit variance, three independent judges each classify every document as ``required'' (essential to answering the query), ``valid'' (relevant but not necessary), or ``invalid'' (does not help answer the query). An ordinal majority vote determines the final classification, with gold-biased tie-breaking: gold documents are kept as ``required'' unless a majority votes them ``invalid,'' while candidate documents need a strict majority to be promoted. If any single run produces a classification consistent with the original gold set, the pipeline short-circuits and retains the original labels.

Only ``required'' documents enter the gold set. ``Valid'' documents are recorded but excluded from the gold set and are not penalized as invalid extra documents during scoring. This prevents near-duplicates and corroborating sources from inflating the invalid count. When the required set changes, the gold answer and answer facts are regenerated from the updated documents, preserving \textbf{anti-hallucination facts} from the original fact set (negative statements designed to catch specific errors). Corrected questions are flagged in the output.

This correction mechanism reflects the reality that gold set quality can improve incrementally as more systems exercise different retrieval strategies against the corpus. It is more feasible to verify that a newly proposed document is relevant than to confirm exhaustively that no better match exists anywhere in the corpus. Over time, as benchmark submissions accumulate, the ground truth converges toward more complete and accurate labels.

\section{Experiments}

\subsection{Baseline Systems Performance}

Three retrieval configurations are evaluated against the full question set, representing the major families of retrieval approaches. \textbf{BM25}~\citep{robertson2009bm25} and \textbf{vector search} are established baselines for keyword and semantic retrieval, respectively. BM25 uses OpenSearch with a standard analyzer over a single concatenated text field. Vector search embeds documents with OpenAI's \texttt{text-embedding-3-large}~\citep{openai2024embeddings} (3072 dimensions) and retrieves via cosine similarity over Qdrant. Both retrieve a fixed top-10 document set per query. \textbf{Bash Agent} retrieval represents a more recent approach following the agentic keyword search paradigm~\citep{subramanian2025keyword}, an approach seeing significant industry adoption.\footnote{{\fontsize{7}{8.5}\selectfont\url{https://vercel.com/changelog/introducing-bash-tool-for-filesystem-based-context-retrieval}}} The agent is equipped with shell tools (\texttt{grep}, \texttt{find}, \texttt{head}, etc.) to iteratively explore the corpus directory structure, selecting a variable number of documents per question with a 10-minute time budget. The Bash Agent was evaluated with low and medium reasoning with no significant difference in the top-level metrics; the low reasoning numbers are reported in Table~\ref{tab:overall_results}. All systems use GPT-5.4~\citep{openai2026gpt54} (Medium Reasoning) for answer generation and answer evaluation to provide a fair comparison.

\begin{table}[t!]
  \caption{Overall performance of baseline retrieval systems. Best result per metric in \textbf{bold}. $\dagger$\,Vector uses OpenAI \texttt{text-embedding-3-large}. $\ddagger$\,Bash Agent uses GPT-5.4 with low reasoning.}
  \label{tab:overall_results}
  \centering
  \begin{tabular}{lccc}
    \toprule
    Metric & BM25 & Vector{\small\textsuperscript{$\dagger$}} & Bash Agent{\small\textsuperscript{$\ddagger$}} \\
    \midrule
    Correctness (\%)       & \textbf{68.8} & 51.4 & 60.6 \\
    Completeness (\%)      & 56.0          & 42.9 & \textbf{61.1} \\
    Document Recall (\%)   & \textbf{68.4} & 46.0 & 55.8 \\
    Invalid Extra Docs     & 9.0           & 9.3  & \textbf{2.0} \\
    \bottomrule
  \end{tabular}
  \\[0.5em]
  \parbox{\textwidth}{\footnotesize \textit{Note:} Correctness and completeness can exceed document recall because (1)~non-gold documents may contain overlapping facts, (2)~partial retrieval can still produce a correct answer, and (3)~some facts are corroborated across multiple gold documents.}
\end{table}

\begin{table}[t!]
  \caption{Per-type correctness (\%) and document recall (\%) for baseline systems. Best result per row in \textbf{bold}. High Level and Info Not Found carry no gold documents, so recall is omitted (---). $\dagger$\,Vector uses OpenAI \texttt{text-embedding-3-large}. $\ddagger$\,Bash Agent uses GPT-5.4 with low reasoning.}
  \label{tab:pertype_results}
  \centering
  \small
  \begin{tabular}{l cc cc cc}
    \toprule
    & \multicolumn{2}{c}{BM25} & \multicolumn{2}{c}{Vector{\small\textsuperscript{$\dagger$}}} & \multicolumn{2}{c}{Bash Agent{\small\textsuperscript{$\ddagger$}}} \\
    \cmidrule(lr){2-3} \cmidrule(lr){4-5} \cmidrule(lr){6-7}
    Question Type & Corr. & Recall & Corr. & Recall & Corr. & Recall \\
    \midrule
    Basic (175)              & \textbf{79.4} & \textbf{77.7} & 50.3 & 48.6 & 65.7 & 61.7 \\
    Semantic (125)           & \textbf{44.8} & \textbf{43.2} & 32.8 & 24.8 & 42.4 & 29.6 \\
    Intra-Doc Reasoning (40) & \textbf{85.0} & \textbf{90.0} & 57.5 & 57.5 & 75.0 & 75.0 \\
    Project Related (40)     & \textbf{60.0} & \textbf{65.5} & 52.5 & 49.2 & 37.5 & 43.2 \\
    Constrained (30)         & \textbf{76.7} & \textbf{85.0} & \textbf{76.7} & 83.3 & \textbf{76.7} & 78.3 \\
    Conflicting Info (20)    & \textbf{90.0} & \textbf{82.5} & 60.0 & 55.0 & 80.0 & 72.5 \\
    Completeness (20)        & \textbf{40.0} & 46.5          & \textbf{40.0} & 33.4 & 35.0 & \textbf{59.0} \\
    Miscellaneous (20)       & 85.0          & 90.0          & 75.0 & 75.0 & \textbf{90.0} & \textbf{100.0} \\
    High Level (10)          & 50.0          & ---           & \textbf{60.0} & --- & \textbf{60.0} & --- \\
    Info Not Found (20)      & \textbf{100.0}& ---           & \textbf{100.0}& --- & \textbf{100.0}& --- \\
    \bottomrule
  \end{tabular}
\end{table}

BM25 leads on correctness and document recall. Vector search underperforms expectations even on semantic questions (32.8\% correctness, 24.8\% recall), the category explicitly designed to favor embedding-based retrieval by limiting keyword overlap. There are a couple of likely contributing factors to the findings. One component may be that embedding models are trained on public corpora (web pages, Wikipedia, forums) and have limited exposure to enterprise-specific vocabulary such as project codenames and internal acronyms, structured formats like tickets and CRM records, and conversational styles typical of internal messaging. Vector retrieval may also be better suited to broad exploration than targeted lookup: cases where the answer is synthesized from related documents rather than pulled from a specific one. Vector search's higher correctness over BM25 on High Level questions is consistent with this, though the category is too small (10 questions, no recall) to weigh heavily in aggregate metrics. This merits further investigation, which we leave to future work. The baseline results do, however, reinforce the premise motivating EnterpriseRAG-Bench: public benchmark performance does not necessarily transfer to enterprise retrieval.

Table~\ref{tab:pertype_results} breaks performance down by question category, showing where each system's overall numbers come from. The Bash Agent is the only system to exceed 60\% overall completeness, achieving the highest score on that metric (61.1\%) despite trailing BM25 on correctness and recall overall. Its strength is iterative discovery: once it reaches a relevant document, it can explore nearby files and similar patterns to find related documents. This possibly explains the high recall on \textit{completeness}-type questions (59.0\% vs.\ 46.5\% for BM25 and 33.4\% for vector), the hardest category overall. It is also the only system to achieve perfect recall on miscellaneous questions, highlighting its strength at retrieving documents outside of dense topic areas. However, the Bash Agent approach comes at a substantial cost in latency and LLM cost: each question may take up to 10 minutes of iterative exploration, compared to a fraction of a second for retrieval plus a few seconds for generation with either BM25 or vector search.

\subsection{Scaling Behavior}
\label{sec:scale}

\begin{figure}[t!]
  \centering
  \includegraphics[width=\textwidth]{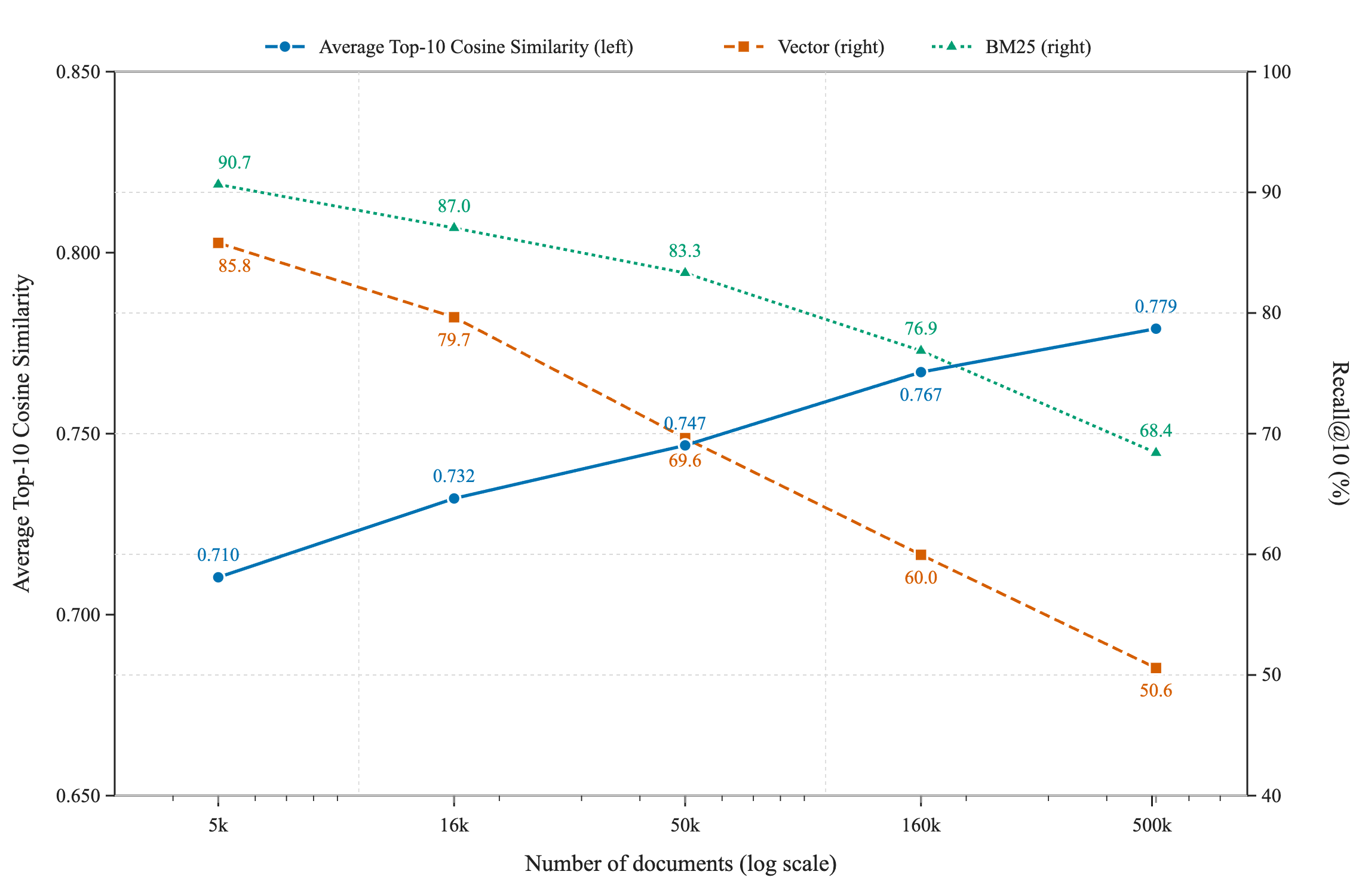}
  \caption{Recall@10 and average cosine similarity of the 10 nearest neighbors vs.\ corpus size. Note: vector search for this experiment uses exact nearest neighbors.}
  \label{fig:recall-acs}
\end{figure}

We also analyze retrieval behavior at different document scales by sampling progressively larger subsets of EnterpriseRAG-Bench. We evaluate at five corpus sizes: 5{,}000, 16{,}000, 50{,}000, 160{,}000, and the full 511{,}962 documents. Figure~\ref{fig:recall-acs} shows the expected pattern: as the corpus grows, top-10 local cosine similarity rises while Recall@10 declines for both BM25 and vector search.

\section{Limitations}
\label{sec:limitations}

The EnterpriseRAG-Bench dataset is synthetic: no real enterprise corpus of comparable scale and diversity can be released publicly without compromising confidentiality. While the generation framework addresses many of the resulting challenges, several limitations remain inherent to LLM-based corpus construction at this scale. We group them into generation fidelity, representation constraints, design tradeoffs, and evaluation.

\paragraph{Generation fidelity.}
Four properties of LLM-based generation reduce the realism of individual documents. First, \textit{long-tail errors} are statistically expected rather than edge cases at this scale: the initial corpus contained documents with control characters in file names, malformed metadata, and other artifacts that required manual removal. Second, \textit{synthetic randomness artifacts} arise because LLMs fall back on recognizable patterns when generating values that should appear arbitrary: Unix timestamps converge on round numbers like \texttt{123456789}, company names default to ``ACME'' variants, and numerical distributions lack the irregularity of real data. These artifacts are cosmetic rather than structural, but could provide unintended retrieval signals. Third, \textit{document structure and path drift} occurs despite guidance from \texttt{agents.md} files: the LLM does not always follow the specified layout, and natural-language-driven generation offers no strict schema enforcement. The same drift applies to file paths: some documents end up in locations that break the expected hierarchy (e.g., GitHub Pull Requests placed directly under the top-level directory rather than inside a repository). At the scale of the released dataset these deviations are a minority, but they are non-negligible. Fourth, \textit{LLM-generated conversations are unrealistically focused}: real Slack threads feature off-topic tangents, misunderstandings, and unrelated interjections that LLM-generated threads rarely reproduce. This affects retrieval difficulty in both directions: there are fewer trivially irrelevant documents to discard, but also less realistic noise for systems to filter through.

\paragraph{Representation constraints.}
All documents are stored as flattened JSON key-value pairs to standardize processing and export. This does not capture the nested structures, embedded objects, or rich formatting that real enterprise documents naturally exhibit: email threads with quoted reply trees, CRM records with nested contact hierarchies, or wiki pages with embedded tables and media.

\paragraph{Design tradeoffs.}
Three deliberate tradeoffs constrain the dataset's scope. First, the high-volume generation flow restricts per-document context to keep costs manageable, with no access to the employee directory and only shallow cross-references scoped to leaf topics. Many high-volume documents consequently reference fictional contributors not grounded in the organizational chart, and cross-document coherence is weaker than in the high-fidelity core. This is benign for the released question set, which draws its multi-document and people-dependent questions exclusively from the high-fidelity portion, but it limits the kinds of questions that can be reliably generated against the bulk of the corpus. Second, the dataset simulates a single technology company at a specific stage and scale. Real enterprises vary widely across industries, organizational maturity, communication culture, and tooling choices, and the released corpus will be more representative of some organizations than others. Third, because several generation steps feed accumulated context back into the prompt, generation quality may degrade at scales substantially beyond the released corpus. The current methodology has not been stress-tested at larger volumes.

\paragraph{Evaluation.}
At 500,000 documents, exhaustive annotation is infeasible. Despite multi-retriever pooling and LLM-based conformance checks, the gold document sets and answers are revisable hypotheses rather than fixed ground truth, as discussed in Section~\ref{sec:evaluation}. The correction mechanism addresses this incrementally, but any individual question's gold labels may be incomplete or improvable, a property shared with virtually all large-scale retrieval benchmarks. We plan to release periodic updates that incorporate corrections surfaced by benchmark participants, so that gold labels improve over time as more systems exercise different retrieval strategies against the corpus.

\section{Conclusion and Future Work}

We present EnterpriseRAG-Bench, a benchmark for evaluating RAG systems over company-internal knowledge. Our contribution consists of a 500,000-document synthetic corpus spanning nine enterprise source types, 500 questions across ten categories, and a correction-aware evaluation framework. We release the dataset, generation framework, evaluation harness, and a public leaderboard.

The generation framework supports adaptation to different industries, company sizes, and source applications without changes to the core methodology, enabling teams to produce tailored enterprise benchmarks rather than relying only on the released corpus.

Several directions for future question types remain open:

\begin{itemize}[leftmargin=2em]
  \item \textbf{Multi-modal retrieval.} Questions that relate to images and other data that is not faithfully presented as text.
  \item \textbf{True multi-hop reasoning.} Questions where the answer to one retrieval step reveals what to search for next, requiring sequential discovery across documents.
  \item \textbf{High-volume aggregation.} Questions requiring a large fraction of the corpus to stress-test exhaustive lookup at scale.
  \item \textbf{Recency-aware retrieval.} Questions asking for the latest document matching loosely specified criteria, combining topical relevance with temporal constraints.
  \item \textbf{People-centric questions.} Questions about employee responsibilities, expertise, and relationships. People are referenced inconsistently across sources (full names, handles, titles) and are rarely the primary subject of any single document.
\end{itemize}

The benchmark is designed to improve over time. The leaderboard uses the initial released gold set to ensure consistent comparison; accumulated corrections from participating systems will be incorporated into future versioned releases. We invite the community to submit systems and contribute corrections that strengthen the ground truth for future participants.

\bibliographystyle{plainnat}
\bibliography{references}

\begin{thebibliography}{18}
\providecommand{\natexlab}[1]{#1}
\providecommand{\url}[1]{\texttt{#1}}
\expandafter\ifx\csname urlstyle\endcsname\relax
  \providecommand{\doi}[1]{doi: #1}\else
  \providecommand{\doi}{doi: \begingroup \urlstyle{rm}\Url}\fi

\bibitem[Bajaj et~al.(2018)Bajaj, Campos, Craswell, Deng, Gao, Liu, Majumder,
  McNamara, Mitra, Nguyen, et~al.]{bajaj2018msmarco}
Payal Bajaj, Daniel Campos, Nick Craswell, Li~Deng, Jianfeng Gao, Xiaodong Liu,
  Rangan Majumder, Andrew McNamara, Bhaskar Mitra, Tri Nguyen, et~al.
\newblock {MS MARCO}: A human generated machine reading comprehension dataset.
\newblock \emph{arXiv preprint arXiv:1611.09268}, 2018.

\bibitem[Chen et~al.(2021)Chen, Chen, Smiley, Shah, Borber, and
  Bendersky]{chen2021finqa}
Zhiyu Chen, Wenhu Chen, Charese Smiley, Sameena Shah, Iana Borber, and Michael
  Bendersky.
\newblock {FinQA}: A dataset of numerical reasoning over financial data.
\newblock In \emph{Proceedings of EMNLP}, 2021.

\bibitem[Chen et~al.(2025)]{chen2025browsecomp}
Zhiyu Chen et~al.
\newblock {BrowseComp-Plus}: A controlled evaluation framework for browsing
  agents.
\newblock \emph{arXiv preprint arXiv:2508.06600}, 2025.

\bibitem[{Databricks}(2025)]{databricks2025karl}
{Databricks}.
\newblock Meet {KARL}: A faster agent for enterprise knowledge, powered by
  custom {RL}.
\newblock Technical report, Databricks, 2025.

\bibitem[Jin et~al.(2019)Jin, Dhingra, Liu, Cohen, and Lu]{jin2019pubmedqa}
Qiao Jin, Bhuwan Dhingra, Zhengping Liu, William Cohen, and Xinghua Lu.
\newblock {PubMedQA}: A dataset for biomedical research question answering.
\newblock In \emph{Proceedings of EMNLP}, 2019.

\bibitem[Kwiatkowski et~al.(2019)Kwiatkowski, Palomaki, Redfield, Collins,
  Parikh, Alberti, Epstein, Polosukhin, Devlin, Lee,
  et~al.]{kwiatkowski2019natural}
Tom Kwiatkowski, Jennimaria Palomaki, Olivia Redfield, Michael Collins, Ankur
  Parikh, Chris Alberti, Danielle Epstein, Illia Polosukhin, Jacob Devlin,
  Kenton Lee, et~al.
\newblock Natural questions: A benchmark for question answering research.
\newblock \emph{Transactions of the Association for Computational Linguistics},
  7:\penalty0 453--466, 2019.

\bibitem[Lloyd(1982)]{lloyd1982kmeans}
Stuart~P. Lloyd.
\newblock Least squares quantization in {PCM}.
\newblock \emph{IEEE Transactions on Information Theory}, 28\penalty0
  (2):\penalty0 129--137, 1982.

\bibitem[Malkov and Yashunin(2020)]{malkov2020hnsw}
Yu~A. Malkov and D.~A. Yashunin.
\newblock Efficient and robust approximate nearest neighbor search using
  hierarchical navigable small world graphs.
\newblock \emph{IEEE Transactions on Pattern Analysis and Machine
  Intelligence}, 42\penalty0 (4):\penalty0 824--836, 2020.

\bibitem[Muennighoff et~al.(2023)Muennighoff, Tazi, Magne, and
  Reimers]{muennighoff2023mteb}
Niklas Muennighoff, Nouamane Tazi, Lo{\"i}c Magne, and Nils Reimers.
\newblock {MTEB}: Massive text embedding benchmark.
\newblock In \emph{Proceedings of EACL}, 2023.

\bibitem[{OpenAI}(2024)]{openai2024embeddings}
{OpenAI}.
\newblock New embedding models and {API} updates.
\newblock OpenAI Blog, 2024.

\bibitem[{OpenAI}(2026)]{openai2026gpt54}
{OpenAI}.
\newblock {GPT-5.4}.
\newblock OpenAI, 2026.

\bibitem[Petroni et~al.(2021)Petroni, Piktus, Fan, Lewis, Yaber,
  et~al.]{petroni2021kilt}
Fabio Petroni, Aleksandra Piktus, Angela Fan, Patrick Lewis, Yacine Yaber,
  et~al.
\newblock {KILT}: A benchmark for knowledge intensive language tasks.
\newblock In \emph{Proceedings of NAACL}, 2021.

\bibitem[Robertson and Zaragoza(2009)]{robertson2009bm25}
Stephen Robertson and Hugo Zaragoza.
\newblock The probabilistic relevance framework: {BM25} and beyond.
\newblock \emph{Foundations and Trends in Information Retrieval}, 3\penalty0
  (4):\penalty0 333--389, 2009.

\bibitem[Subramanian et~al.(2025)]{subramanian2025keyword}
Sathya Subramanian et~al.
\newblock Keyword search is all you need.
\newblock \emph{arXiv preprint arXiv:2602.23368}, 2025.

\bibitem[Thakur et~al.(2021)Thakur, Reimers, R{\"u}ckl{\'e}, Srivastava, and
  Gurevych]{thakur2021beir}
Nandan Thakur, Nils Reimers, Andreas R{\"u}ckl{\'e}, Abhishek Srivastava, and
  Iryna Gurevych.
\newblock {BEIR}: A heterogeneous benchmark for zero-shot evaluation of
  information retrieval models.
\newblock In \emph{Proceedings of NeurIPS}, 2021.

\bibitem[Trivedi et~al.(2022)Trivedi, Balasubramanian, Khot, and
  Sabharwal]{trivedi2022musique}
Harsh Trivedi, Niranjan Balasubramanian, Tushar Khot, and Ashish Sabharwal.
\newblock {MuSiQue}: Multihop questions via single hop question composition.
\newblock \emph{Transactions of the Association for Computational Linguistics},
  10:\penalty0 539--554, 2022.

\bibitem[van~der Maaten and Hinton(2008)]{vandermaaten2008tsne}
Laurens van~der Maaten and Geoffrey Hinton.
\newblock Visualizing data using {t-SNE}.
\newblock \emph{Journal of Machine Learning Research}, 9:\penalty0 2579--2605,
  2008.

\bibitem[Yang et~al.(2018)Yang, Qi, Zhang, Bengio, Cohen, Salakhutdinov, and
  Manning]{yang2018hotpotqa}
Zhilin Yang, Peng Qi, Saizheng Zhang, Yoshua Bengio, William Cohen, Ruslan
  Salakhutdinov, and Christopher~D Manning.
\newblock {HotpotQA}: A dataset for diverse, explainable multi-hop question
  answering.
\newblock In \emph{Proceedings of EMNLP}, 2018.

\end{thebibliography}

\newpage
\appendix

\begin{center}
\Large\textbf{Appendices}
\end{center}
\vspace{1em}

\section{Repository and Code Guide}

The generation framework and evaluation harness are available in the repository at \url{https://github.com/onyx-dot-app/EnterpriseRAG-Bench}. The dataset itself is distributed as release artifacts (not checked into the repository) and can be downloaded from the latest GitHub release.\footnote{\url{https://github.com/onyx-dot-app/EnterpriseRAG-Bench/releases/latest}}

\subsection{Dataset Artifacts}

The release includes the following artifacts:

\begin{itemize}[leftmargin=2em]
  \item \texttt{all\_documents.zip} --- The full document corpus as a single archive.
  \item Per-source slices (\texttt{<source\_type>\_slice\_<N>.zip}) --- Individual archives of up to 5,000 documents each, for partial downloads.
  \item \texttt{questions.jsonl} --- The 500 benchmark questions. Also available at the repository root.
  \item \texttt{extra\_questions.jsonl} --- 100 additional metadata-dependent questions excluded from the core benchmark.
\end{itemize}

\subsection{Evaluation Harness}

Retrieval systems are expected to produce an answers file in JSONL format where each line contains a \texttt{question\_id}, an \texttt{answer}, and a list of \texttt{document\_ids} for the retrieved documents.

\begin{lstlisting}
{"question_id": "qst_0001", "answer": "Your answer text...",
 "document_ids": ["dsid_abc", "dsid_def"]}
\end{lstlisting}

\textbf{Metrics-based evaluation} scores a single system on the four metrics described in Section~\ref{sec:metrics}:

\begin{lstlisting}
python -m src.scripts.answer_evaluation.metrics_based_eval \
    --answers-file answer_evaluation/answers.jsonl
\end{lstlisting}

\textbf{Comparative evaluation} scores two systems head-to-head with the three-judge consensus process:

\begin{lstlisting}
python -m src.scripts.answer_evaluation.comparative_eval \
    --answer-file-1 answer_evaluation/system_1.jsonl \
    --answer-file-2 answer_evaluation/system_2.jsonl
\end{lstlisting}

Questions that receive gold set corrections during evaluation are flagged in the results and written to a separate updated questions file.

\subsection{Generation Pipeline}

The generation framework lives under \texttt{src/} and is organized into four stages, each containing numbered scripts that are run sequentially (Table~\ref{tab:pipeline}).

\begin{table}[t!]
  \caption{Generation pipeline stages and their purpose.}
  \label{tab:pipeline}
  \centering
  \small
  \begin{tabular}{clcp{6.5cm}}
    \toprule
    Stage & Path & Scripts & Purpose \\
    \midrule
    1 & \texttt{data\_gen\_stage\_1\_...} & 1--9 & Scaffolding, high-fidelity documents, completeness clusters, volume documents \\
    2 & \texttt{data\_gen\_stage\_2\_...} & 1--4 & Random shuffle, LLM-based shuffle, miscellaneous files, near-duplicates \\
    3 & \texttt{data\_gen\_stage\_3\_...} & 1--11 & Ten question types plus optional metadata questions \\
    4 & \texttt{data\_gen\_stage\_4\_...} & 1    & Export and packaging \\
    \bottomrule
  \end{tabular}
\end{table}

All scripts are resumable: they check for existing output and prompt before regenerating. Intermediate state is persisted in \texttt{generation\_cache/}, allowing long-running stages to be interrupted and resumed. The repository's \texttt{quickstart.md} provides full invocation commands with recommended flags for every step.

\section{Document and Noise Generation Details}
\label{app:docgen}

\subsection{Project-Level Scaffolding}
\label{app:project_scaffolding}

High-level initiatives are decomposed into projects---smaller efforts spanning engineering, go-to-market, operations, and internal functions---each targeting on the order of 100 documents. Projects are individually enriched: the LLM generates a detailed description and a manifest of documents to produce, using tool access (glob, tree) to inspect the existing directory structure and \texttt{agents.md} files so that proposed documents respect source-type constraints. Each project is further annotated with relevant employees from the directory, including their roles within the project.

\subsection{High-Fidelity Generation}
\label{app:highfidelity}

For each project, every document is generated with access to: (a)~its own file name, path, and description; (b)~the company overview; (c)~the full enriched project description including the manifest of all sibling documents; (d)~the \texttt{agents.md} files along its path; and (e)~a read tool that allows the model to inspect previously generated project documents when additional detail is needed.

Completeness clusters are generated through a separate step that produces small groups of 4--10 documents where every document in the cluster is loaded into the model's context simultaneously. Unlike the project-level generation, which operates over ${\sim}100$ documents and pulls in siblings on demand, completeness clusters guarantee full mutual visibility. Each cluster is anchored to a target question so that the generation knows which facts must be distributed across the group, ensuring that no single document contains the complete answer.

\subsection{High-Volume Generation and Topic Scaffolding}

For each source type, the system generates a hierarchical decomposition of topics and subtopics, with estimated document counts per topic calibrated to match expected real-world distributions (e.g., an engineering Slack workspace should cover the full range of discussions one would actually expect, not just the handful of themes an LLM gravitates toward). Topics are recursively split until each leaf contains at most 500 documents. During generation, the model sees only the company overview, key initiatives, the \texttt{agents.md} for the target source type, and the file paths of other documents within the same leaf topic. This narrow context window, compared to the full project scaffolding used in high-fidelity generation, substantially reduces cost per document while the leaf-level path visibility nudges the model to produce complementary rather than overlapping content.

This design involves deliberate tradeoffs. Documents in the high-volume flow are not grounded in the employee directory, so they reference fictional contributors not present in the organizational chart. Cross-references are shallow and scoped to leaf topics rather than enforced globally. These limitations are benign for the released question set, which draws its multi-document and people-dependent questions from the high-fidelity core.

\subsection{Noise Generation Procedures}
\label{app:noise_procedures}

\paragraph{Random shuffle.}
A specified percentage of documents (5\% in the released dataset) are randomly relocated within their source type. Documents are selected via a random walk over the directory tree so that selection reflects directory structure rather than being skewed by per-folder file counts. Cross-source shuffling is excluded to preserve format consistency: a ticketing system document with fields like ``assignee'' and ``closed date'' would be incoherent if placed in a Slack channel.

\paragraph{LLM-based shuffle.}
Real misfiling is not uniformly random; documents tend to end up in structurally plausible but incorrect locations: adjacent directories, parent/child folders, or nearby branches of the hierarchy. An additional 3\% of documents (excluding those already shuffled) are relocated using an LLM that receives the document's original path, its contents, and the source's directory structure, and proposes a destination that is reasonable but suboptimal. Cumulatively, 8\% of documents in the released dataset are affected by some shuffle operation.

\paragraph{Near-duplicates with conflicting facts.}
We sample documents using the directory-walk procedure described above and generate a near-duplicate at a new location with specific facts changed. Unlike the shuffle operations, near-duplicates may cross source-type boundaries: a Slack conversation might contain an update on information originally recorded in a CRM record. The generation model receives the original document's path and contents along with the \texttt{agents.md} files at the new location, producing a document that is structurally appropriate for its destination while introducing controlled factual divergence from the original. These pairs are tracked in a generation cache and later used to produce conflicting-info questions.

\paragraph{Miscellaneous directories and files.}
Real archives contain informal discussion, work-in-progress drafts, and ad-hoc notes in poorly organized locations. We introduce miscellaneous directories (e.g., \texttt{slack/memes}, \texttt{google\_drive/.../misc-assets}, \texttt{github/hackathons}) through a human-in-the-loop process: given the existing directory structure, the LLM proposes candidate directories that the user reviews and finalizes. These directories are populated with informal, off-topic documents generated with awareness of the company overview, relevant \texttt{agents.md} files, and previously generated miscellaneous files (to avoid the same topic-clustering problem that motivated the scaffolding approach). The volume of miscellaneous files is intentionally small: they exist not to shift the corpus distribution but to support a question category that targets informal content stored in less predictable locations.

\section{Question Generation Flows}
\label{app:questions}

This appendix details the per-type generation procedures. The family-level overview is in Section~\ref{sec:questions}.

\subsection{Single-Document Flows}

\textbf{Basic}, \textbf{Semantic}, \textbf{Intra-Document Reasoning}, and \textbf{Miscellaneous} questions each begin by sampling a single document and generating one question from it. All four follow the same pipeline---generate question, validate conformance, produce gold answer, extract atomic facts---but differ in sampling criteria, prompt guidance, and validation logic.

\paragraph{Basic.}
A random document is sampled and the LLM generates a retrieval-oriented question. The prompt steers toward moderate keyword overlap: enough that the question is clearly related to the document, but not so much that exact phrase matching trivially solves it. Key guidance includes avoiding long verbatim phrases from the source, avoiding complex multi-part questions, and using varied phrasing so that the question set does not collapse into a uniform ``What is\ldots'' distribution. Few-shot examples of good and bad questions are provided to calibrate length and specificity.

\paragraph{Semantic.}
The generation pipeline is identical to basic questions, but the prompt guidance is substantially different. Where basic questions permit moderate keyword overlap, semantic questions actively suppress it: the LLM is instructed to avoid keyword matches where possible, limit qualifiers and scoping details, and produce ``challenging, loose-match, semantic-type'' queries. The few-shot examples are correspondingly harder, demonstrating roundabout phrasing that relies on meaning rather than surface terms.

\paragraph{Intra-Document Reasoning.}
This type targets a specific failure mode: systems that chunk documents and embed each chunk independently tend to miss questions whose answer requires combining information from distant sections of the same document. At the same time, expanding every retrieved document in full risks flooding the context window with irrelevant text. Intra-doc reasoning questions probe this tension directly. They also screen for a subtler failure: cases where one section of a document appears to contain the answer but a different section disqualifies it entirely.

During sampling, a minimum length threshold filters out documents too short to support cross-section questions. The LLM is guided to generate questions that require relating information from different parts of the document. A validation step then checks that the question cannot be answered from any single contiguous passage; questions that fail this check are rejected and regenerated.

\paragraph{Miscellaneous.}
These questions target the informal, off-topic files introduced during noise generation (e.g., files in \texttt{slack/memes} or \texttt{github/hackathons}). These documents sit outside the main scaffolding-driven generation and present a distinct retrieval challenge: they are topically peripheral and stored in less predictable locations, so systems tuned to the corpus's dominant themes may overlook them. The generation flow is identical to basic questions---same prompt guidance, same validation---but restricted to the set of miscellaneous documents tracked in the generation cache.

\subsection{Multi-Document and Corpus-Exploration Flows}

\textbf{Project Related}, \textbf{Constrained}, \textbf{High Level}, and \textbf{Info Not Found} questions require the generation model to explore the corpus before formulating a question. Each flow equips the LLM with file-system tools and provides context designed to guide exploration toward productive regions of the corpus.

\paragraph{Project Related.}
The model receives a project's full document manifest and a read tool, then iteratively reads documents within the project until it identifies a question requiring information from multiple sources. The prompt steers the model toward four patterns:

\begin{itemize}[leftmargin=2em]
  \item \textbf{Multi-document synthesis}: the answer requires combining parts from multiple documents to build a coherent picture.
  \item \textbf{Cross-cutting themes}: items mentioned incidentally across multiple documents but never the primary focus of any single one.
  \item \textbf{Contradictions}: conflicting information across documents, requiring the answer to acknowledge and reconcile the discrepancy.
  \item \textbf{Causal chains}: tracing a cause-and-effect relationship through multiple documents to construct a coherent narrative.
\end{itemize}

All documents read during exploration are tracked. After question generation, a separate step receives the question and all read documents and identifies the minimal subset necessary for a correct answer; this filtered set becomes the gold documents. Because these questions are broader than single-document types, there is a higher likelihood that documents outside the original project also contain relevant information, an inherent property of multi-document questions at this corpus scale that the quality assurance process is designed to address.

\paragraph{Constrained.}
A constrained question contains qualifiers that narrow the correct answer to a single document, even though many other documents in the corpus are superficially relevant. The qualifiers act as filters: each one eliminates documents that would otherwise match. This tests whether a retrieval system can distinguish surface-level relevance from true relevance under specific conditions: naive systems will return many partially matching documents, but only the document satisfying all constraints contains the correct answer.

Generation proceeds in two phases. In the exploration phase, an LLM agent equipped with file-system tools (glob, grep, read, etc.) performs the following:

\begin{enumerate}[leftmargin=2em]
  \item Explores the directory structure to find clusters of topically related documents.
  \item Reads documents within a cluster and identifies the axes along which they differ.
  \item Formulates a question whose qualifiers narrow the answer to a single gold document, and identifies explicit distractor documents.
\end{enumerate}

Previously explored document paths from prior constrained question runs are provided to prevent over-clustering in the same corpus regions. In the second phase, the gathered context is turned into a gold answer and facts. Unlike single-document types where facts are extracted from a separately produced answer, constrained questions generate the answer and facts jointly. The distractor documents are used to produce anti-hallucination facts: negative statements crafted to catch specific errors a system might make if it retrieves a distractor instead of the gold document (e.g., ``The answer must not claim the Postgres database is the in-cluster StatefulSet, since the scenario specifies externally managed Postgres'').

\paragraph{High Level.}
These questions are answerable from the company overview and initiatives---the top-level scaffolding documents---but should not be answerable from any single document in the corpus. They test whether a system can synthesize broad organizational knowledge distributed across many documents. Generation proceeds in three stages:

\begin{enumerate}[leftmargin=2em]
  \item \textbf{Candidate generation.} The LLM receives the company overview and initiatives and produces a batch of candidate queries. Guidance steers questions toward cross-cutting organizational patterns rather than point lookups, and requires varied phrasing and topic spread across the batch.
  \item \textbf{Validation.} Each candidate is checked by a separate LLM agent equipped with file-system tools (glob, grep, read, etc.). The agent attempts to find a single document that directly answers the query; if it succeeds, the query is rejected as too specific.
  \item \textbf{Answer and fact generation.} For each validated query, a gold answer is produced from the company overview and initiatives, and answer facts are extracted.
\end{enumerate}

High-level questions carry no expected document identifiers, since the answer derives from organizational context distributed across the corpus and multiple different document subsets may each suffice. There is no strict guarantee that every question is fully answerable from the document set. The assumption is that across the full set of high-level questions, the relevant organizational context will typically exist somewhere in the corpus.

\paragraph{Info Not Found.}
An LLM agent equipped with file-system tools (glob, grep, read, etc.) explores the corpus to find a cluster of topically related documents. After reading several documents in the cluster, the agent identifies the dimensions along which they differ and crafts a natural-sounding query that is related to the cluster's subject matter but not answerable from the documents. Corpus exploration is essential: by reading what the corpus actually contains, the model can deliberately adjust scoping, constraints, or details to fall just outside the available information, producing questions that are reliably unanswerable rather than generating blind queries. The query is designed so that surface-level keyword or topic matching would retrieve plausible-looking documents, but any answer derived from them would be hallucinated.

Previously explored document paths are tracked in a generation cache and provided in subsequent runs to prevent the agent from revisiting the same corpus regions. The gold answer and evaluation fact are predefined rather than generated: the expected answer states that the information is not available, and the single fact checks that the response acknowledges this rather than fabricating an answer.

\subsection{Generation-Dependent Flows}

\textbf{Conflicting Info} and \textbf{Completeness} questions depend on document structures created during earlier pipeline stages rather than discovering them through exploration.

\paragraph{Conflicting Info.}
These questions target cases where one document supersedes another, typically when a newer document invalidates specific facts from an older one. This creates both a reasoning and a retrieval challenge: if the retriever returns only the superseded document, the system will report outdated information as fact. The generation model receives both documents from the near-duplicate pairs generated during the noise stage (Appendix~\ref{app:noise_procedures}) and produces a question that requires reconciling the discrepancy: identifying which document supersedes the other and providing the correct, up-to-date answer. As with constrained questions, the answer and facts are generated jointly, and anti-hallucination facts are included to catch responses that report only the outdated information.

\paragraph{Completeness.}
These questions test whether a system can exhaustively retrieve all documents needed to answer a question, not just a relevant subset. Partial retrieval is insufficient: if even one required document is missing, aggregated counts will be wrong, lists will be incomplete, and comparisons will be skewed, making completeness a direct test of recall depth.

Questions are built from the tightly coupled document clusters generated during the high-fidelity phase (Appendix~\ref{app:highfidelity}), where 4--10 documents were produced with full mutual visibility and facts deliberately distributed so that no single document contains the complete answer. Because the target question already exists from the cluster generation step, this flow focuses on filtering: an LLM evaluates each document's contribution and removes any that prove unnecessary in practice. Clusters that reduce to fewer than two required documents after filtering are discarded. The remaining set defines the gold documents, and answer facts are extracted as individually verifiable claims to enable fine-grained recall measurement.

\section{Dataset Statistics}

This appendix provides exact corpus and question-level statistics for the released dataset.

\subsection{Corpus Composition}

Table~\ref{tab:corpus} provides exact document counts and length statistics by source type.

\begin{table}[t!]
  \caption{Corpus composition by source type. Lengths are measured in characters from the exported plain-text representation, sampled over 500 documents per source type.}
  \label{tab:corpus}
  \centering
  \small
  \begin{tabular}{clrrrr}
    \toprule
    \# & Source Type   & Doc Count & Mean Length & Median Length & P95 Length \\
    \midrule
    1  & Slack         & 285,605 &  3,427 &  3,294 &  5,618 \\
    2  & Gmail         & 121,390 &  7,101 &  6,909 &  9,390 \\
    3  & Linear        &  35,308 &  5,059 &  4,875 &  7,529 \\
    4  & Google Drive  &  25,108 &  7,160 &  7,094 & 10,113 \\
    5  & HubSpot       &  15,017 &  3,141 &  3,004 &  4,756 \\
    6  & Fireflies     &  10,173 & 11,605 & 11,074 & 16,315 \\
    7  & GitHub        &   8,052 &  4,741 &  4,664 &  6,910 \\
    8  & Jira          &   6,120 &  5,563 &  5,596 &  7,454 \\
    9  & Confluence    &   5,189 &  9,852 &  9,660 & 13,834 \\
    \midrule
       & \textbf{Total} & \textbf{511,962} & & & \\
    \bottomrule
  \end{tabular}
\end{table}

The variation is substantial: meeting transcripts (Fireflies) and wiki pages (Confluence) are 3--4$\times$ longer than Slack threads and CRM records (HubSpot), reflecting the different information density and formatting conventions of each source type.

Slack threads are notably longer than typical real-world threads, where many are short or even single-message. This is discussed further in Section~\ref{sec:limitations}.

\subsection{Source Type Coverage Across Question Types}

Table~\ref{tab:coverage} shows how often each source type appears as a gold document across question categories. A single question may span multiple source types, so row totals can exceed the question count.

\begin{table}[t!]
  \caption{Number of gold documents per source type for each question category. High Level and Info Not Found carry no gold document identifiers by design. A single question may span multiple source types.}
  \label{tab:coverage}
  \centering
  \footnotesize
  \setlength{\tabcolsep}{4pt}
  \begin{tabular}{l*{9}{r}r}
    \toprule
    Question Type & Confl. & Fire. & GH & Gmail & GDrive & HS & Jira & Linear & Slack & Total Qs \\
    \midrule
    Basic              & 19 & 11 & 22 & 15 & 24 & 18 & 21 & 24 & 21 & 175 \\
    Semantic           & 15 &  4 & 14 & 18 & 14 & 14 & 23 & 10 & 13 & 125 \\
    Intra-Doc Reas.    &  2 &  3 &  2 &  9 &  0 &  1 &  2 &  9 & 12 &  40 \\
    Project Related    & 30 &  0 & 18 &  7 &  7 &  0 & 29 &  9 & 14 &  40 \\
    Constrained        & 16 &  0 &  0 &  0 &  1 &  0 & 13 &  2 &  3 &  30 \\
    Conflicting Info   & 14 &  2 &  2 &  2 &  8 &  1 &  4 &  4 &  1 &  20 \\
    Completeness       & 16 &  3 &  1 &  3 &  2 &  0 &  5 &  0 &  5 &  20 \\
    Miscellaneous      &  3 &  2 &  1 &  0 &  3 &  0 &  2 &  0 &  9 &  20 \\
    High Level         & \multicolumn{9}{c}{---} & 10 \\
    Info Not Found     & \multicolumn{9}{c}{---} & 20 \\
    \bottomrule
  \end{tabular}
\end{table}

Two patterns stand out. Confluence and Jira are substantially overrepresented as gold sources relative to their corpus share (1.0\% and 1.2\% of documents, respectively, but appearing in 24\% and 21\% of questions with gold documents). This reflects the nature of structured documentation: wiki pages and tickets are more likely to contain definitive, self-contained answers than conversational messages. Conversely, Slack constitutes 56\% of the corpus but appears as a gold source in only 17\% of questions.

\subsection{Gold Document Set Size by Question Type}

Table~\ref{tab:goldsize} reports the number of gold documents per question. Single-document types (Basic, Semantic, Intra-Document Reasoning, Miscellaneous) have exactly one gold document by construction. Conflicting Info always pairs two documents. Constrained questions have 1--2 gold documents, as the count is determined by the LLM based on query complexity. Project Related and Completeness questions require the most documents, with Completeness reaching up to 10.

\begin{table}[t!]
  \caption{Gold document set size statistics by question type. High Level and Info Not Found carry no expected documents.}
  \label{tab:goldsize}
  \centering
  \begin{tabular}{lrrrr}
    \toprule
    Question Type & Mean & Median & Min & Max \\
    \midrule
    Basic                    & 1.0 & 1   & 1 & 1  \\
    Semantic                 & 1.0 & 1   & 1 & 1  \\
    Intra-Document Reasoning & 1.0 & 1   & 1 & 1  \\
    Miscellaneous            & 1.0 & 1   & 1 & 1  \\
    Conflicting Info         & 2.0 & 2   & 2 & 2  \\
    Constrained              & 1.4 & 1   & 1 & 2  \\
    Project Related          & 4.2 & 3.5 & 2 & 9  \\
    Completeness             & 6.5 & 6   & 2 & 10 \\
    High Level               & --- & --- & --- & --- \\
    Info Not Found           & --- & --- & --- & --- \\
    \bottomrule
  \end{tabular}
\end{table}

\end{document}